\documentclass[aps,pra,reprint,superscriptaddress]{revtex4-1}

\usepackage{amsmath}
\usepackage{graphicx}
\usepackage{epsfig}

\begin{document}

\title{High-precision measurements and theoretical calculations \\ of indium excited-state polarizabilities}

\author{N. B. Vilas}
\affiliation{Department of Physics, Williams College, Williamstown, MA 01267}
\author{B.-Y. Wang}
\affiliation{Department of Physics, Williams College, Williamstown, MA 01267}
\author{P. M. Rupasinghe}\altaffiliation{Current address: Dept. of Physics, SUNY Oswego, Oswego, NY 13126}
\affiliation{Department of Physics, Williams College, Williamstown, MA 01267}
\author{D. L. Maser}
\affiliation{Department of Physics, Williams College, Williamstown, MA 01267}
\author{M. S. Safronova}
\affiliation{Department of Physics and Astronomy, University of Delaware, Newark, DE 19716}
\affiliation{Joint Quantum Institute, National Institute of Standards and Technology and the University of Maryland, Gaithersburg, MD 20742}
\author{U.~I.~Safronova}
\affiliation{Department of Physics, University of Nevada, Reno, NV 89557}
\author{P. K. Majumder}
\email{pmajumde@williams.edu}
\affiliation{Department of Physics, Williams College, Williamstown, MA 01267}

\date{\today}

\begin{abstract}
We report measurements of the $^{115}$In $7p_{1/2}$ and $7p_{3/2}$ scalar and tensor polarizabilities using two-step diode laser spectroscopy in an atomic beam. The scalar polarizabilities are one to two orders of magnitude larger than for lower lying indium states due to the close proximity of the $7p$ and $6d$ states. For the scalar polarizabilities, we find values (in atomic units) of $1.811(4) \times 10^5$ $a_0^3$ and $2.876(6) \times 10^5$ $a_0^3$ for the $7p_{1/2}$ and $7p_{3/2}$ states respectively. We estimate the smaller tensor polarizability component of the $7p_{3/2}$ state to be $-1.43(18) \times 10^4$ $a_0^3$. These measurements represent the first high-precision benchmarks of transition properties of such high excited states of trivalent atomic systems. We also present new \textit{ab initio} calculations of these quantities and other In polarizabilities using two high-precision relativistic methods to make a global comparison of the accuracies of the two approaches. The precision of the experiment is sufficient to differentiate between the two theoretical methods as well as to allow precise determination of the indium $7p-6d$ matrix elements. The results obtained in this work are applicable to other heavier and more complicated systems, and provide much needed guidance for the development of even more precise theoretical approaches.
\end{abstract}


\maketitle

\section{Introduction}

 Accurate knowledge of atomic properties has been critical for many applications, including the search for physics beyond the Standard Model \cite{SafBudDem17}, time and frequency metrology \cite{NicCamHut15,BelHinPhil14}, the suppression of decoherence in quantum information processing \cite{ZhaRobSaf11,GolNorKol15}, degenerate quantum gases \cite{PorSafDer14}, and many others.
Progress in the development of high-precision theory \cite{PorBelDer09,SafronovaCIAll2009,TupKozSaf16} has yielded accurate predictions of many needed properties while high-precision measurements, such as \cite{BarStaLem08,NicCamHut15,KeeHanWoo11,MidFalLis12,BelSheLem12,Indium2013,Indium2013}, have provided experimental benchmarks for the refinement and improvement of theory. Further progress in atomic theory is needed for the design and interpretation of experiments, the development of concepts for next-generation experiments and precision measurement techniques, and the quantification and reduction of uncertainties and decoherence. For example, recent proposals for the development of clocks and tests of fundamental physics with highly charged ions \cite{BerDzuFla10,SafDzuFla14} have highlighted the urgent need for new, more precise theoretical predictions in these systems. Further development of theory requires associated improvement in precision measurements to serve as accurate experimental benchmarks. Such measurements in alkali and alkaline-earth metal atoms have been indispensable for the development of current theoretical approaches and the understanding of their uncertainties. Precise measurements in more complicated atomic systems are far scarcer, and urgently needed.

Trivalent group IIIA atoms like indium and thallium have long been considered promising experimental testbeds in the search for discrete symmetry violations and other quantities of fundamental physical interest, such as permanent electric dipole moments (EDMs) \cite{FortsonPNC1995,DeMilleTlEDM2002,IndiumEDM2011, SafronovaTlEDM2012}. In-like and Tl-like ions are also excellent candidates for the development of ultra-precision clocks and the search for the variation of the fine-structure constant $\alpha$. Despite very high ionization energies, certain highly charged ions have transitions that lie in the optical range due to level crossing and are very sensitive to $\alpha$-variation \cite{BerDzuFla12}. In-like and Tl-like ions are particularly well suited for
the experimental search for such transitions \cite{SafDzuFla14}, with Tl-like Cf$^{17+}$ appearing to be a particularly attractive
candidate \cite{BerDzuFla12}. While techniques for the \emph{ab initio} atomic theory work necessary to interpret such experiments are well-developed for single-valence alkali systems, theoretical methods for the treatment of trivalent systems have only more recently demonstrated significant improvements in precision \cite{Safronova2013,SafronovaCIAll2009, SafronovaIn2007,SafronovaTl2006,SahooDasIn2011}. The theory used to interpret a 1995 measurement of parity nonconservation (PNC) in thallium \cite{FortsonPNC1995, TlPNCTheory2001}, for instance, lags the experimental precision by a factor of three; similar PNC work in cesium does not suffer from similar theoretical limitations \cite{PorBelDer09,DzuBerFla12}. While experimental data are not available for most Tl-like and In-like {\em ionic} systems, we can gain insight to these systems by carefully comparing theory and experiment in neutral group IIIA systems.

Many of the applications listed above require precise knowledge of excited-state atomic properties for which very few experimental benchmarks (beyond frequency interval measurements) exist. The determination of transition matrix elements between excited states is a particularly difficult challenge for both theory and experiment. Measurements of dynamic polarizabilities to provide such benchmarks for divalent systems have recently been proposed \cite{Sr}. The present work supplies new benchmarks for trivalent systems using indium as a test case.

Recently, a 2013 measurement of the polarizability of the $6s_{1/2}$ state in indium \cite{Indium2013} inspired a new series
of calculations using two different high-precision approaches \cite{Safronova2013}. A subsequent measurement of the $6p_{1/2}$ scalar polarizability \cite{Indium2016} in 2016 was sufficiently precise to not only provide a test of theory, but to distinguish between the two slightly different theoretical predictions. In the present work, we present new precision measurements of the indium $7p_{1/2}$ scalar and $7p_{3/2}$ scalar and tensor polarizabilities alongside new \textit{ab initio} calculations of these quantities using two high-precision relativistic methods. We then make a global comparison of the accuracies of the two theoretical approaches using all available experimental data. As will be discussed, the precision of the experiments is sufficient to clearly differentiate between the two theoretical methods. We note that, due to the presence of very nearby $6d$ levels, the scalar polarizabilities of these $7p$ states are 30-50 times larger than those previously measured in our laboratory. As discussed below, these measurements can serve as unambiguous determinations of the $6d-7p$ matrix elements themselves.

\section{Atomic Structure Details}\label{sec:AtomicStructure}

\begin{figure}
\includegraphics[width=.8\columnwidth]{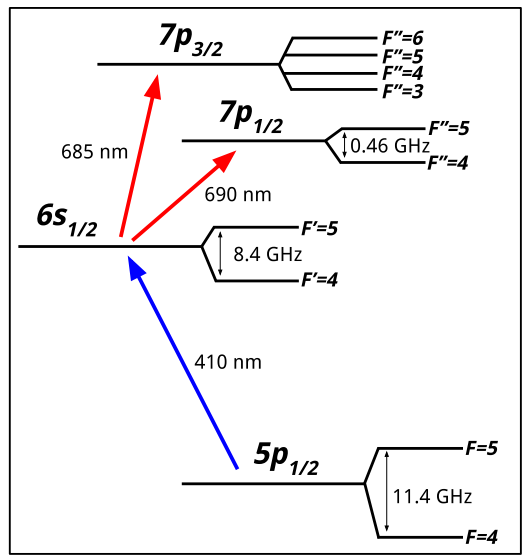}
\caption{Level structure of $^{115}$In states relevant to the present measurements. Our two-step spectroscopic scheme requires that we overlap 410 nm and 690 (685) nm lasers through an atom source to drive the `first-step' $5p_{1/2}-6s_{1/2}$ transition and the `second-step' $6s_{1/2}-7p_{1/2(3/2)}$ transitions.}
\label{fig:IndiumStructure}
\end{figure}

Indium has atomic number $Z=49$ and a ground-state electron configuration given by [Kr]$4d^{10}5s^{2}5p$. This state has electronic angular momentum $J=1/2$, and we notate it as the $5p_{1/2}$ state. In the present experiment we consider three resonance lines. One, at 410 nm, excites the $5p_{1/2} - 6s_{1/2}$ transition, and the other two, at 690 and 685 nm, respectively, are resonant with the excited $6s_{1/2} - 7p_{1/2,3/2}$ transitions. See Fig. \ref{fig:IndiumStructure} for the relevant energy-level structure.

In all of our studies, we focus on the $^{115}\text{In}$ isotope ($96\%$ abundant). Small peaks from $^{113}\text{In}$ are either unresolved or spectroscopically separated. These small features can be explicitly accounted for, but their presence does not contribute in any significant way to our experimental uncertainties. As discussed in Refs. \cite{Indium2013, Indium2016}, $^{115}\text{In}$ has nuclear spin $I=9/2$, meaning that all $J=1/2$ states studied have hyperfine levels $F=4$ and $F=5$, while the $7p_{3/2}$ state has $F=3,4,5,6$. In particular, the $5p_{1/2}$ and $6s_{1/2}$ states considered below have hyperfine splittings (HFS) of 11.4 and 8.4 GHz, respectively, while the various hyperfine splittings for the $7p$ states range from 100 to 500 MHz. 

For the case of our $7p_{1/2}$ Stark shift measurement, since $J=1/2$, there is only a common scalar Stark shift for all hyperfine sublevels, leading to a scalar polarizability, $\alpha_0$, and no tensor component. We expect to observe an energy shift for each sublevel of $\Delta E = -\frac{1}{2}\alpha_0 \mathcal{E}^2$, where $\mathcal{E}$ is the magnitude of the applied electric field, taken to be along the $z$ axis. As discussed in Ref. \cite{Indium2016}, if we keep the first-step excitation laser tuned to the \emph{Stark-shifted} $5p_{1/2} \rightarrow 6s_{1/2}$ resonance, the observed frequency shift in the second-step ($6s_{1/2}\rightarrow 7p_{1/2}$) transition will be exactly given by $\Delta \nu = - \frac{1}{2 h} \left[\alpha_0(7p_{1/2})-\alpha_0(6s_{1/2})\right]\mathcal{E}^2\equiv k_0 \mathcal{E}^2$, where $k_0$ is the scalar Stark shift constant. Using our previous measurement of the $\alpha_0(6s_{1/2}) - \alpha_0(5p_{1/2})$ polarizability difference \cite{Indium2013} in conjunction with theoretical predictions for the very small $\alpha_0(5p_{1/2})$ \cite{Safronova2013}, we can determine a precise value for the $7p_{1/2}$ scalar polarizability, with negligible introduction of additional uncertainties.

\begin{figure}
\includegraphics[width=0.95\columnwidth]{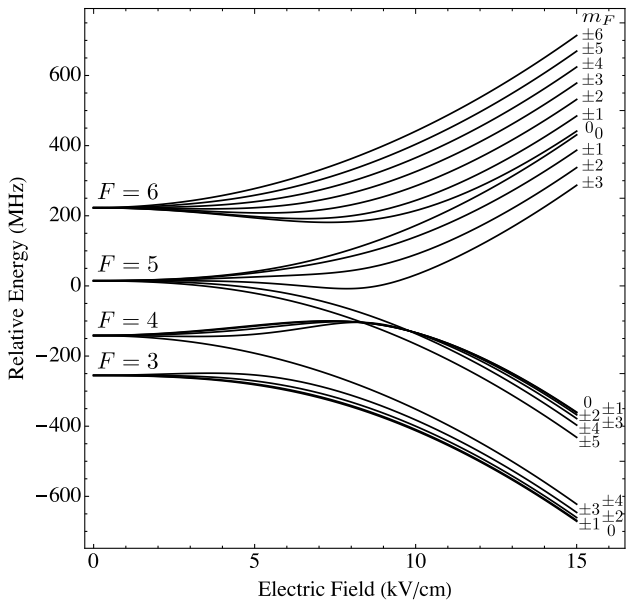} 
\caption{Energy eigenvalue structure under applied electric field for all hyperfine sub-levels of the indium $7p_{3/2}$ state. The $m_F$ designations of the sub-levels are indicated at the far right of the figure. Here, for clarity, we have subtracted out the large scalar shift, $-\frac{1}{2}\alpha_0\mathcal{E}^2$, shared by all $|F,m_F\rangle$ levels.}
\label{fig:TensorLevels}
\end{figure}

In contrast to this, the $7p_{3/2}$ state admits a tensor polarizability in addition to the scalar component discussed above. The tensor component mixes $F$ states; in this case, the Hamiltonian in the presence of an electric field is
\begin{equation}
H = V_S + V_\text{hf}
\label{eqn:7p32Hamiltonian}
\end{equation}
where the hyperfine Hamiltonian, $V_\text{hf}$, can be found in, for instance, Ref. \cite{Gerginov}. The Stark Hamiltonian, $V_S$, is given by
\begin{align}
\langle Fm_F | V_S | F'm_F\rangle = &-\frac{1}{2}\alpha_0 \mathcal{E}^2 \delta_{FF'} \nonumber \\
& - \frac{1}{2}\alpha_2\mathcal{E}^2\langle Fm_F | Q | F'm_F \rangle \text{,}
\end{align}
where the hyperfine-basis Stark mixing operator $Q$ is derived in Ref. \cite{Schmieder1972}. The Hamiltonian is block diagonal in $m_F$ because we take the electric field along the quantization axis. We also note that this result is \emph{not} perturbative, as the Stark shift is of the same order as the hyperfine structure in this state. Figure \ref{fig:TensorLevels} shows the field-dependent results of a numerical diagonalization of the full Hamiltonian for a range of electric fields attainable in the laboratory. In this figure we have omitted the large, common shift of all levels due to the scalar polarizability for clarity.

In analogy to the Stark shift constant used in the extraction of the $7p_{1/2}$ scalar polarizability above, we can introduce an `effective' Stark shift constant for each particular sub-level of the $7p_{3/2}$ state,
\begin{equation}
k_\text{eff} = k_0 + c(F,m_F)k_2
\label{eqn:keff}
\end{equation}
that combines the effects of the tensor and scalar polarizabilities so that the observed shift of a level $|Fm_F\rangle$ between fields $\mathcal{E}_1$ and $\mathcal{E}_2$ is given by $\Delta \nu = k_\text{eff} (\mathcal{E}_2^2 - \mathcal{E}_1^2)$. Here $k_0 = -\frac{1}{2h}[\alpha_0(7p_{3/2}) - \alpha_0(6s_{1/2})]$ as above, and $k_2 = -\frac{1}{2h}\alpha_2(7p_{3/2})$ analogously. Given this parameterization, we note that both $k_\text{eff}$ and $k_0$ have negative values, whereas $k_2$ itself turns out to be positive (though much smaller in magnitude). The coefficients $c(F,m_F)$ are level-dependent factors of order unity that reflect the relative shift of distinct hyperfine states and can be calculated numerically by diagonalizing the Hamiltonian in Eq. \ref{eqn:7p32Hamiltonian}. The sign of these coefficients is positive for the group of upward-trending states in Fig. \ref{fig:TensorLevels} and negative for the lower frequency, downward-trending states.

Such a formulation is only approximate, as the shift effected by the tensor polarizability is not purely quadratic in the electric field. Equivalently, one can view the coefficient $c(F,m_F)$ as having a slight electric-field dependence. Nonetheless, for the limited range of large electric fields used to extract the tensor polarizability, the uncertainty in a measurement of $\alpha_2$ due to imprecision in this simple field-independent model for $c(F,m_F)$ is at the level of 0.5\% or below, and can be neglected when compared to other experimental errors, as discussed below. The final fractional experimental uncertainty in the tensor polarizability of the $7p_{3/2}$ state is quite large in comparison to our scalar polarizability measurements, due both to its size relative to the scalar component, as well as to the complications of composite spectral peaks associated with multiple unresolved, but non-degenerate magnetic sublevels. We note that our final $\sim 12\%$ experimental uncertainty in this quantity is in agreement with, and of comparable precision to, the theoretical prediction presented below.

\section{Experimental Details}\label{sec:ExpDetails}

\subsection{Atom beam source and electric field production}

We perform polarizability measurements in a collimated beam of indium atoms to which precisely calibrated DC electric fields are applied in order to effect a static Stark shift. The portion of apparatus used for atom and electric field production is practically identical to that described in Refs. \cite{Indium2013, Indium2016}. In brief, the atomic beam is contained in a home-built vacuum chamber held at approximately $10^{-7}$ Torr through the use of two diffusion pumps. A sample of indium metal is first heated in a molybdenum crucible to roughly $1100^\circ$C. Several collimating stages are then applied along a $\sim 0.5$-meter beam path between the source oven and the interaction region. Due to this geometrical collimation, when we direct the 410 nm first-step laser transversely to the atomic beam, we see a residual Doppler width of roughly 100 MHz.

In the measurement region, the atomic beam passes between two circular, 10-cm diameter stainless steel capacitor plates whose separation was measured to be 1.0038(5) cm. We apply voltages of up to 20 kV using a commercial high-voltage (HV) supply \footnote{Glassman ER40P07.5} and measure them using a high-precision voltage divider and a calibrated $6\frac{1}{2}$ digit voltmeter \footnote{Keithley 2100} in parallel with the field plates. We direct the second-step red laser in counter-propagating fashion to the blue beam and the lasers interact with the indium atoms over a 2-cm-wide region in the center of the field plates. Three orthogonal sets of magnetic field coils cancel the Earth's field to roughly 1 $\mu$T in the measurement region.

\subsection{Optical setup}\label{sec:OpticalSetup}

\begin{figure}
\includegraphics[width=0.45\textwidth]{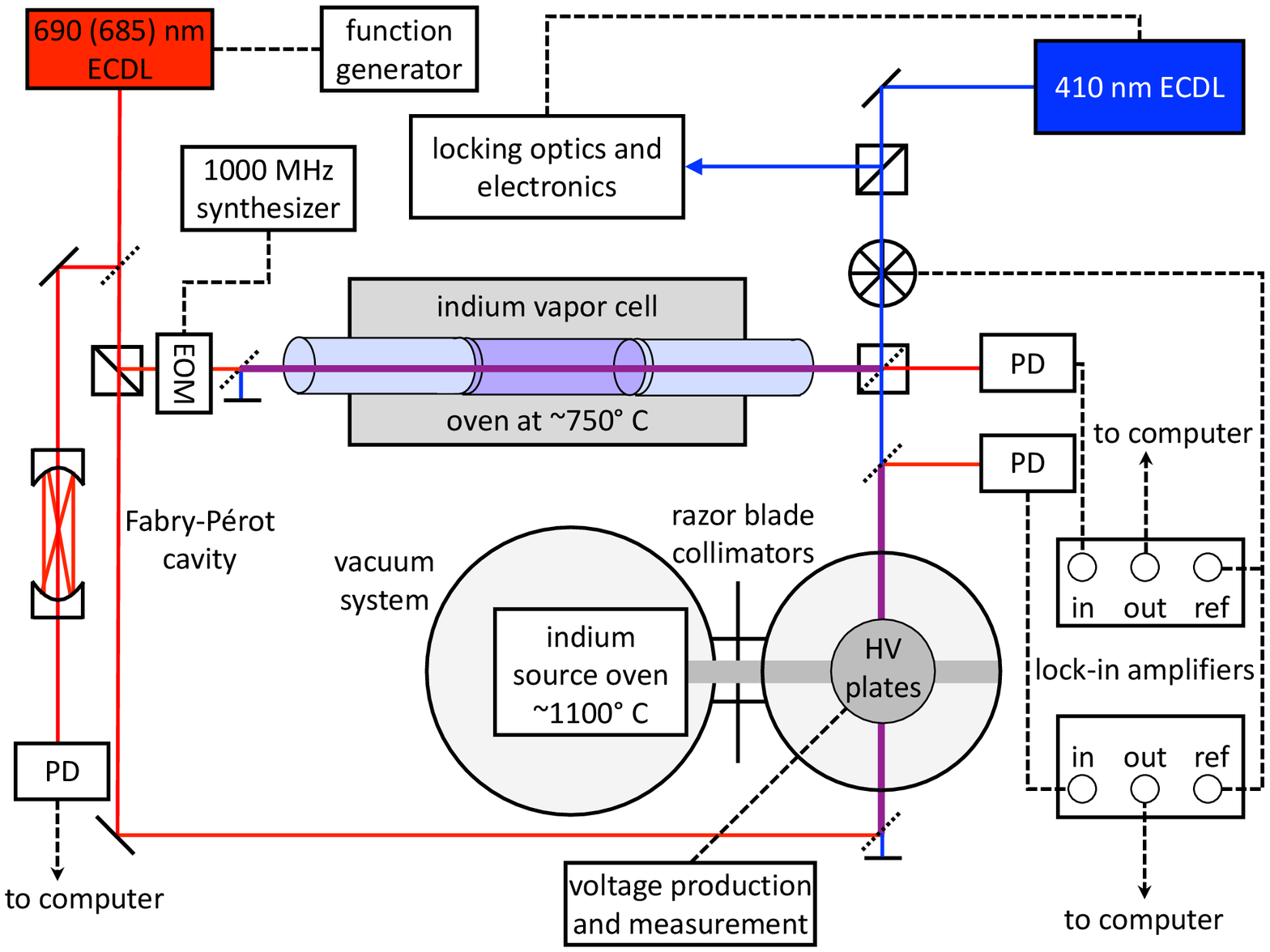}
\caption{Simplified diagram of the full optical setup used in indium polarizability measurements. Here PD refers to a photodetector, EOM to an electro-optic modulator, and ECDL to an external cavity diode laser. Two acousto-optic modulators (not shown) are inserted in the path of the blue laser beam component which is incident on the atomic beam apparatus in order to maintain resonance for the first-step transition as the electric field is changed in the interaction region (see text). }
\label{fig:ExpSetupAll}
\end{figure}

Our experiment makes use of a two-step laser spectroscopy technique similar to that described in Refs. \cite{Thallium2014, Indium2016}. We use two external cavity diode lasers (ECDLs) in the Littrow configuration. The first ECDL (Toptica DL 100) is locked to one of the Doppler-broadened 410 nm $5p_{1/2}(F=4,5)\rightarrow 6s_{1/2}(F'=4,5)$ first-step hyperfine transitions in a supplementary, field-free, heated vapor cell using an FM spectroscopy technique described in Ref. \cite{Indium2016}. This technique allows us to achieve frequency stabilization of better than 1 MHz RMS over timescales of several hours. A setup consisting of two acousto-optic modulators (AOMs) is used to shift the frequency of the 410 nm light directed to the atomic beam to remain resonant with the Stark-shifted first-step transition there. The precise frequency shift required for a given electric field is well-known from our previous measurement \cite{Indium2013}.

A second, home-built ECDL is directed in a spatially overlapping, counter-propagating geometry through the atomic beam, and is scanned over the hyperfine levels of the relevant 690 (685) nm $6s_{1/2}\rightarrow7p_{1/2(3/2)}$ second-step transition. To observe the very weak red laser absorption signal, we modulate the 410 nm light directed through the atomic beam with an optical chopping wheel at $\sim 1$ kHz. We then detect the red absorption with a 10 MHz-bandwidth photodiode and demodulate at the first-step chopping frequency using a lock-in amplifier. This serves to eliminate background and greatly improve the signal-to-noise ratio for this second-step signal. Because the locked 410 nm laser only interacts with a limited range of atomic velocities, this technique produces a virtually Doppler-free second-step spectrum. Despite low optical depths in the atomic beam ($\sim 10^{-3}$ for the 410 nm transition) and relatively small line strengths associated with the indium $6s - 7p$ transitions (1 to 2 orders of magnitude weaker than in the case of our recent $6s-6p$ polarizability work), the lock-in detection scheme is sufficiently sensitive to yield second-step hyperfine spectra with peaks resolvable at the 1-2 MHz level for a typical 10-s scan.

Using the same detection scheme as in the atom beam, we separately monitor the second-step hyperfine spectra in a heated, field-free vapor cell. The resulting high-resolution spectra (see, for example, the lower plot in Figs. \ref{fig:7p12data} and \ref{fig:7p32data}) serve as stable frequency references from which to measure Stark shifts in the atomic beam. Additionally, the red light directed to the cell is modulated at $\omega_m = 2\pi \times 1000$ MHz using an electro-optic modulator (EOM) -- by doing so, we introduce first-order sidebands at $\pm \omega_m$ into the vapor cell spectra, which are used to calibrate the frequency axes of our scans. Finally, a small portion of red laser light is directed into a Fabry-P\'erot cavity (free spectral range $\approx 363$ MHz) to aid in frequency axis linearization during analysis. Fig. \ref{fig:ExpSetupAll} shows a simplified diagram of the complete optical setup.

\subsection{Data acquisition procedure}

We use a LabVIEW program to control and measure the applied electric field in the atomic beam unit, apply the proper AOM frequency to maintain 410 nm resonance with the Stark-shifted transition, and collect Fabry-P\'{e}rot, vapor cell, and atomic beam data for successive laser scans. These are separated into upscans and downscans corresponding, respectively, to increasing and decreasing laser frequency with time.

For $7p_{1/2}$ scalar polarizability measurements, we collect data at electric fields between 1 and 6 kV/cm, alternating between scans with the electric field on and the electric field off. Given the large polarizability of this excited state, this produces readily measurable Stark shifts of order several hundred MHz. We follow a similar procedure for $7p_{3/2}$ scalar polarizability measurements, though here we only collect data for fields up to 3 kV/cm, since for larger fields, the tensor component of the polarizability begins to noticeably complicate the lineshape as can be seen in Fig. \ref{fig:TensorLevels} (potential systematic errors introduced by the tensor polarizability are discussed below). As a means of testing for systematics related to long-term drifts in the electric field, we successively alternate the order in which field-on and field-off scans are collected. Pairs of scans taken in `off $\rightarrow$ on' order are then compared with those taken in the `on $\rightarrow$ off' sequence, as discussed in \cite{Indium2016}.

To measure the $7p_{3/2}$ tensor polarizability, we instead collect data at higher fields near 15 kV/cm. This requires a detuning of the 685 nm laser by roughly 8 GHz from the field-free resonance, meaning that we cannot reference to a (field-free) vapor cell signal. Rather than alternate with field-free data in this configuration, we instead measure the relative Stark shift between different high-field scans, so that the observed frequency shift $\Delta \nu$ between fields $\mathcal{E}_1$ and $\mathcal{E}_2$ is given by $\Delta \nu = k_\text{eff}(\mathcal{E}_2^2-\mathcal{E}_1^2)$ as described in section \ref{sec:AtomicStructure}. A typical collection run acquires scans in the following order: 14 kV/cm $\rightarrow$ 15 kV/cm $\rightarrow$ 16 kV/cm $\rightarrow$ 16 kV/cm $\rightarrow$ 15 kV/cm $\rightarrow$ 14 kV/cm. This allows for the comparison of consecutive scans with increasing vs. decreasing electric fields, a useful check on systematics relating to long-term drifts in the apparatus.

\section{Experimental analysis and results}\label{sec:Results}

\begin{figure}
\includegraphics[width=0.45\textwidth]{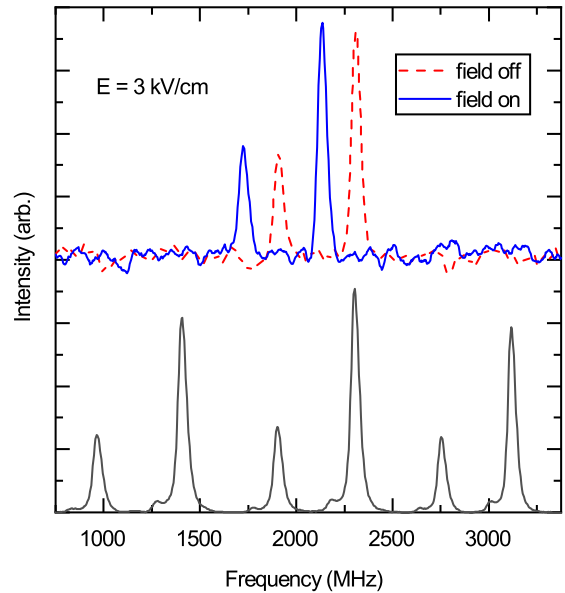}
\caption{Atomic beam spectra for the case of the $(F^\prime=4) \rightarrow (F^{\prime\prime}=4,5)$ transitions of the 690 nm $6s_{1/2} - 7p_{1/2}$ line. The field-off spectrum (red dashed line) and the spectrum with a 3 kV/cm electric field applied (blue solid line) are shown. Displayed below is the corresponding (field-free) vapor cell spectrum, including 1000 MHz FM sidebands, used for frequency referencing and calibration. The small spectral features on the shoulders of the large vapor cell peaks are due to the $^{113}$In isotope (4\% abundance) which we account for in our line shape analysis. As noted in the text, the data shown here represent the average of thirty consecutive field off/field on scan pairs.}
\label{fig:7p12data}
\end{figure}

\begin{figure}
\includegraphics[width=0.45\textwidth]{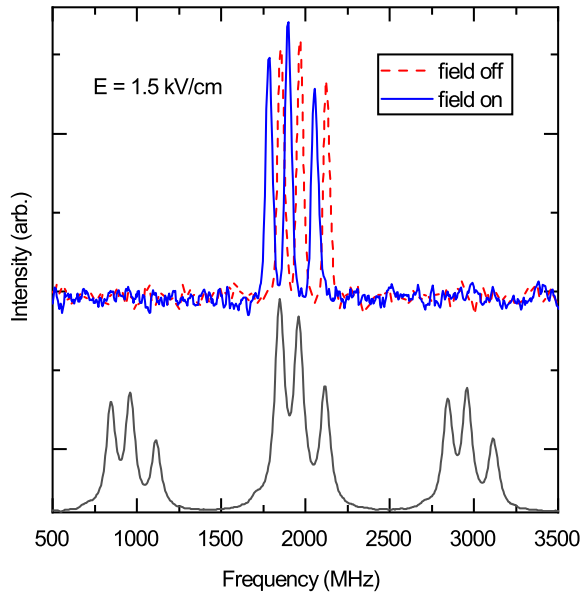}
\caption{Atomic beam spectra for the case of the $(F^\prime=4) \rightarrow (F^{\prime\prime}=3,4,5)$ transitions of the 685 nm $6s_{1/2} - 7p_{3/2}$ line. The field off spectrum (red dashed line) as well as the spectrum with a 1.5 kV/cm electric field applied (blue solid line) are shown. Displayed below is the corresponding (field-free) vapor cell spectrum, including 1000 MHz FM sidebands, used for frequency referencing and calibration. As in the previous figure, the data shown here represent the average of thirty consecutive field off / field on scan pairs.}
\label{fig:7p32data}
\end{figure}

Over the course of several months, several thousand individual field-off / field-on pairs of red laser scans were collected for each of the $6s-7p$ transitions. Over the course of these measurements, in addition to the electric field value, we varied experimental parameters such as the choice of intermediate ($6s$) state hyperfine level, relative optical power and laser polarization, atomic beam source temperature, as well as laser sweep speed and frequency range. Fig. \ref{fig:7p12data} shows typical atomic beam field-off/field red laser spectra for the $7p_{1/2}$ state (top), with the accompanying vapor cell reference/calibration scan below. Fig. \ref{fig:7p32data} shows a similar set of scans for one set of three $7p_{3/2}$ hyperfine sublevels. In both cases, the frequency axes have been linearized and calibrated as noted below and as outlined in detail in Refs. \cite{Indium2013, Indium2016}. We extract Stark shifts for each pair of scans of consecutive field-on / field-off scans; however, for display purposes, in the figures included here, we have averaged the data from 30 consecutive scan pairs taken under identical conditions over the course of roughly 20 minutes.

\subsection{Data analysis procedure}

We extract polarizabilities from collected data following a procedure similar to that described in Ref. \cite{Indium2016}. We first linearize the frequency axes for every scan using the positions of the red Fabry-P\'{e}rot transmission peaks. We then fit field-free vapor cell data to sums of six (nine) Lorentzian peaks, corresponding to two (three) hyperfine peaks and four (six) first-order EOM sidebands at $\pm 1000$ MHz for the $7p_{1/2}$ ($7p_{3/2}$) state. The frequency axis is calibrated by extracting the observed splittings between hyperfine peaks and their corresponding first-order EOM sidebands -- the axis is then scaled to bring these splittings to their known value of 1000 MHz. We then determine the change in the relative position of the atomic beam spectrum and a reference peak from the calibrated vapor cell spectrum upon application of the electric field to determine the Stark shift. For $7p_{3/2}$ tensor polarizability scans, which contain no vapor cell signal, we use the frequency calibration from vapor cell data separately taken both immediately before and after these runs. 

For $7p_{1/2}$ and $7p_{3/2}$ scalar polarizability data, Stark shifts are extracted from atomic beam spectra using two complementary methods. The first (the `Lorentzian method') requires that we fit atomic beam data to sums of two (three) Lorentzians, corresponding to the relevant hyperfine peaks. We then compare resonance locations for field-on and field-off scans. The second (the `overlap method') assumes no functional form and instead computes the sum of squared differences between field-on and field-off scans for a variety of shifts (frequency-axis translations) of the field-on scan. When this value is minimized, the peaks are optimally `overlapped,' and the Stark shift can be determined. The potential line shape systematic errors to which these two methods are susceptible are quite different, so that agreement in the respective results (as we observe) is a good indication of the absence of significant systematics of this type.

For $7p_{3/2}$ tensor polarizability data, taken at higher fields near 15 kV/cm, we observe a spectrum consisting of two well-defined, though composite, peaks (each consisting of several nondegenerate $m_F$ levels) and an unresolved `plateau'-like feature at higher frequency (refer to the high-field region of Fig. \ref{fig:TensorLevels}). By measuring the shift of the lower two peaks between pairs of voltages near 15 kV/cm -- using either of the Lorentzian or overlap methods with a chosen Fabry-P\'{e}rot peak as a stable frequency reference -- we can extract a value for $k_\text{eff}$ in equation \ref{eqn:keff} above. When combined with the value of $k_0$ derived from low-field measurements, this yields a value for $k_2$ and therefore $\alpha_2$. Independently, we confirmed that the transmission peaks of our passively-stabilized Fabry-P\'{e}rot, with its low-expansion-material construction, drift by no more than a few MHz over time scales of one hour. Given the Stark shift differences that we measure at these fields (roughly 500 MHz), any drift-related errors are negligible compared to our final tensor polarizability experimental uncertainty.

\subsection{Error analysis and final results}

Our general approach to investigation of systematic errors, which we follow in this work, is thoroughly discussed in Refs. \cite{Indium2013, Indium2016}. Our results for the $^{115}$In $7p_{1/2}$ and $7p_{3/2}$ Stark shift constants $k_0$ and $k_2$, along with relevant statistical and systematic error budgets, are presented in Table \ref{tab:ErrorBudgets}. We determine statistical uncertainties by both assembling histograms of all data, and also considering the weighted average of runs at a given high voltage value taken over a number of different runs and days. Our final statistical error reflects this observed scatter in these sets of data runs taken at all electric field values. We also create and analyze histograms of subsets of data, such as shown in Fig. \ref{fig:7p12analysis}a, for the the $6s-7p_{1/2}$ 690 nm transition, where the Stark shift constant for all $\sim 400$ field-off/field-on scan pairs for E = 3 kV/cm are plotted. Our various statistical approaches produce final average values for data subsets that are in very good agreement.

 \subsubsection{Scalar polarizabilities}
We first bisect the data in various ways based on laser sweep direction, intermediate hyperfine level, order of field-off / field-on sequencing, spectral peak analysis method, etc., and look for statistically significant differences. Occasionally, among some data subsets, these comparisons yield small resolved differences, at the level of 1.5 to 2 (combined) standard deviations, in which case we include associated contributions to the total error budget in Table \ref{tab:ErrorBudgets}.
 
\begin{figure}
\includegraphics[width=0.4\textwidth]{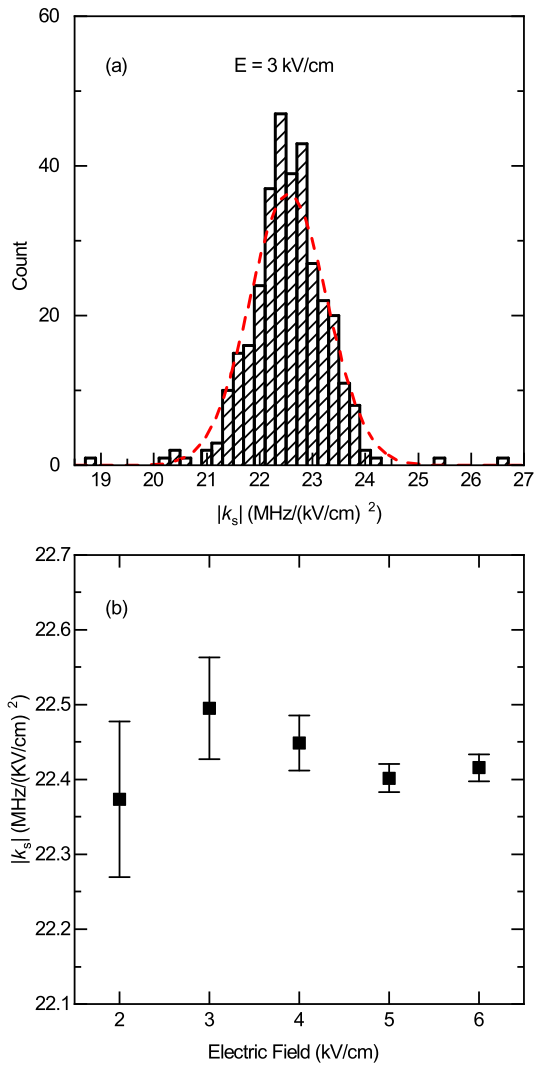}
\caption{(a) For the 690 nm $6s-7p_{1/2}$ transition, we plot the Stark shift constant derived from roughly 400 field-off / field-on scan pairs taken with E = 3 kV/cm. A Gaussian curve is laid over the data for display purposes. Central values and standard errors from such analyses complement a weighted average analysis approach to arrive at final statistical averages and uncertainties. (b) All $7p_{1/2}$ Stark shift data, with Stark shift constant plotted versus electric field to explore potential field-dependent systematic errors. An analysis of these data shows the absence of a statistically resolved correlation. }
\label{fig:7p12analysis}
\end{figure}

We also consider potential systematic errors by searching for correlations of measured polarizabilities with, for example, electric field value and laser power. An example of this is shown in Fig. \ref{fig:7p12analysis}b, where all of our $6s-7p_{1/2}$ Stark shift constant results have been plotted vs. electric field. While, as expected, the {\em precision} of the polarizability determination is much greater at larger field (where the much larger Stark shift can be measured with much greater fractional accuracy), we see no resolved trend in the central values as the field is varied. Also considered are error contributions from imprecision in the measurement and calibration of the applied electric field, due to uncertainty in the field plate separation as well as the applied voltage. Errors due to the calibration and linearization of the frequency axis are also quantified by fitting Fabry-P\'{e}rot and vapor cell spectra using a variety of different methods.

\begin{figure}
\includegraphics[width=0.4\textwidth]{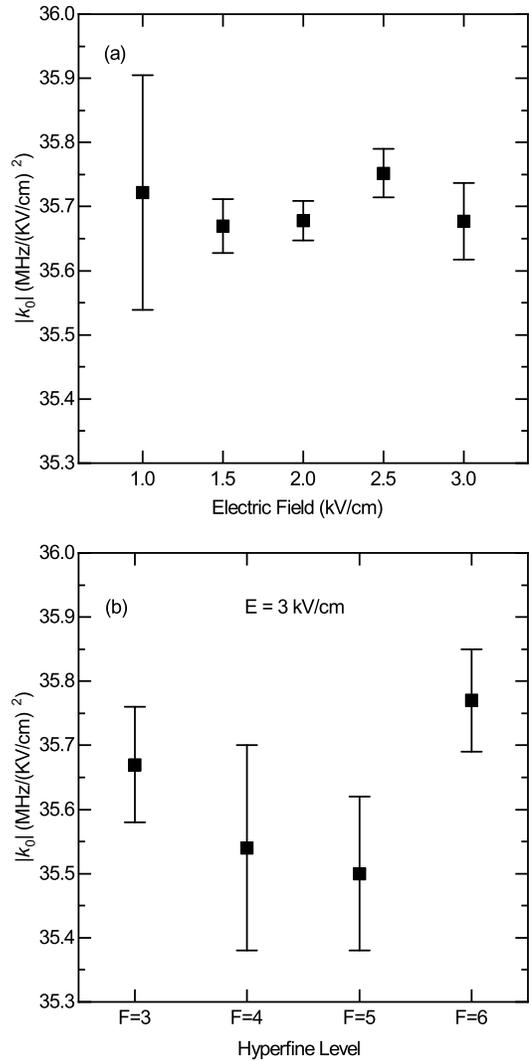}
\caption{(a) For the case of all data collected for the transition path $6s_{1/2}(F^\prime = 5)$ to $7p_{3/2}(F^{\prime\prime} = 4,5,6)$, we average the results of the Stark shift constant for each upper-level hyperfine state, and the resulting average Stark shift constants at each electric field value are plotted. Given that the tensor polarizability is expected to contribute at higher fields, it is notable that over the 1 to 3 kV/cm range shown here we see no statistically significant variation in the measured polarizability. (b) At the highest field used, where we expect some hyperfine peak broadening and possible peak asymmetry, we investigate the dependence of the measured Stark shift constant on hyperfine level, now including data taken for each of the four $7p_{3/2}$-state sub levels.}
\label{fig:7p32analysis}
\end{figure}

In the case of the $7p_{3/2}$ state, we can only access three of the four hyperfine levels in a given two-step excitation path, due to selection rules. We have collected data for both the 3-4-5 hyperfine spectra and the 4-5-6 spectra. It is particularly important to study potential field-dependent systematics here, since we know that the tensor component of the polarizability will eventually cause broadening of hyperfine peaks, and differential Stark shift rates as the electric field increases and tensor contributions to the polarizability become significant. Fig. \ref{fig:7p32analysis}a shows the polarizability determinations averaged over hyperfine levels in all of our 4-5-6 spectra for fields between 1 and 3 kV/cm, a range over which we expect the tensor contributions to be negligible. Fig. \ref{fig:7p32analysis}b shows the 3 kV/cm subset of the data for the $7p_{3/2}$ scalar polarizability measurement, where we have plotted the Stark shift constant for each hyperfine level individually, now including all four of the hyperfine levels. At this largest field value, we would expect any possible systematic error introduced by tensor polarizability-induced hyperfine line broadening and potential line shape asymmetry to be most noticeable. Similar analyses at all field values, while in some cases revealing variation across hyperfine levels that is slightly in excess of the intrinsic statistical uncertainties (for which we include an additional `hyperfine level dependence' systematic error), show no evidence of the type of tensor polarizability trends predicted for higher fields in Fig. \ref{fig:TensorLevels} .Furthermore, as can be seen in that figure, taking the average of all hyperfine Stark shifts at low fields should make us even more immune to any residual tensor effects.

In all cases, contributions from systematic errors remain below the 0.5\% level. Varying laser polarizations will also potentially affect spectra peak determination due to changing rates of excitation for nondegenerate, unresolved $m_F$ levels contained in each observed peak, and we have been careful to explore a variety of polarizations for both lasers in our data sets. In Table \ref{tab:ErrorBudgets} we have included small contributions from this and all other systematic errors that we have considered.
 
\subsubsection{$7p_{3/2}$ tensor polarizability}
\begin{figure}
\includegraphics[width=0.4\textwidth]{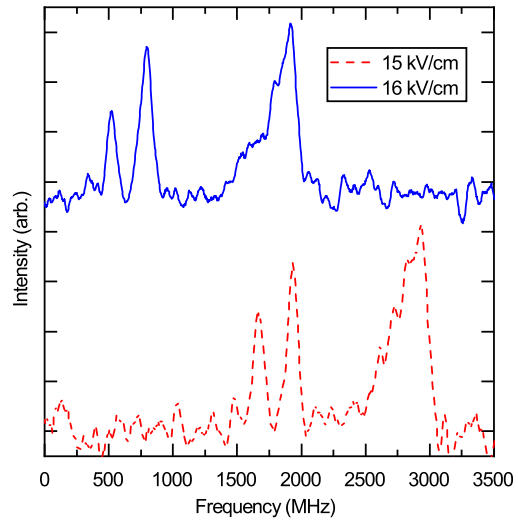}
\caption{For the 685 nm $6s-7p_{3/2}$ transition, spectra are shown for the case of E = 15 kV/cm and 16 kV/cm, showing a roughly 1 GHz overall Stark shift. For each electric field value, we have averaged a series of scans taken consecutively over a period of 10 minutes for display purposes. Referring to Fig. \ref{fig:TensorLevels}, one can see experimental evidence supporting our numerical model, where two relatively sharp experimental peaks are accompanied by a broad composite feature at the higher frequency end of both of the scans shown. }
\label{fig:7p32highfield}
\end{figure}

Having extracted a reliable value for the $7p_{3/2}$ scalar polarizability from the low-field data, we are able to analyze the high-field data to infer a value for the tensor component. As laid out in section II, we first extract $k_{\text{eff}}$ from a pair of two different high field scans, focusing in particular on the shift in the two lower-frequency (composite) spectral peaks, such as can be seen in Fig. \ref{fig:7p32highfield}. Numerical modeling allows us to estimate a range of c$(F, m_{F})$ coefficient values for the set of magnetic sub levels contained within each composite peak. Since we cannot predict the exact weightings of the components within the composite peak, we assign a systematic composite line shape error as part of the analysis. By considering all pairs of 14-15-16 kV/cm data scans, and subtracting the known scalar Stark shift coefficient, we can obtain a final extracted value for $k_2$, and hence $\alpha_2$.
 
Varying relative laser polarization significantly affects these high-field spectra, since the excitation probabilities among the various $7p_{3/2}(F,m_F$) sub levels is highly sensitive to the polarization selection rules. We have collected high-field spectra for several choices of polarization. Within the final experimental uncertainty which we quote, we see consistent results across various choices of polarization values. The tensor component for the polarizability for this indium state has the opposite sign from the scalar component, and is more than an order of magnitude smaller. Both because of its relative size, and the line shape complications alluded to here, our estimate for $\alpha_2$ has a final fractional uncertainty of roughly 12\%. Though this precision is far poorer than all of our recent scalar polarizability measurements, our experimental uncertainty is comparable to the estimated theory uncertainty for the tensor component, and, given these respective uncertainties, is in good agreement with that prediction (see Table \ref{tab}).

\begin{table*}
\begin{tabular}{l | c | c | c}
\hline \hline
& $k_0(7p_{1/2}-6s_{1/2})$ & $k_0(7p_{3/2}-6s_{1/2})$ & $k_2(7p_{3/2})$ \\[3pt]
\hline
\textbf{Result [MHz (kV/cm)$\mathbf{^{-2}}$]} & -22.402 & -35.646 & +1.78 \\[3pt]
\hline
\textbf{Statistical Error} & 0.021 & 0.036 & 0.09\\[3pt]
\textbf{Systematic Error Sources} & & & \\
Laser scan direction & 0.011 & 0.010& 0.05 \\
Frequency calibration & 0.004 & 0.005& 0.04 \\
Scan linearization & 0.002 & 0.003 & 0.003 \\
Electric field calibration & 0.022 & 0.036 & 0.08 \\
Laser power and polarization & 0.005 & 0.006 & 0.07 \\
First-step hyperfine transition & 0.033 & --- & --- \\
Fitting method & 0.018 & 0.011 & --- \\
Hyperfine level dependence & --- & 0.050 & 0.14 \\
Unresolved sub-levels, composite peaks & --- & --- & 0.08 \\[3pt]
\hline
\textbf{Combined Error Total} & 0.050 & 0.076 & 0.23 \\
\hline \hline
\end{tabular}
\caption{Final experimental results, with statistical and systematic error contributions, for the Stark shift constants $k_0$ and $k_2$ of the $6s_{1/2}-7p_{1/2,3/2}$ transitions in $^{115}\text{In}$.}
\label{tab:ErrorBudgets}
\end{table*}

\section{Theory}

Indium can be treated as a system with one valence electron and [$1s^22s^22p^63s^23p^63d^{10}4s^24p^64d^{10}5s^2$]
core or as a trivalent system with the $5s^2$ open shell. In the first case, one can use the method developed to treat alkali-metal atoms where single, double, and partial triple excitations (LCCSDpT) of the Dirac-Fock wave function are included to all orders \cite{SafJoh08}. We refer to this method as CC for brevity in the text and tables below. The advantage of this approach is a more complete inclusion of the correction to the dipole operator, described, for example in a review \cite{SafJoh08}. The disadvantage is the inability to explicitly treat three particle states, such as $5s5p^2$, which appear very low in the In spectrum and lie near the $5s^27s$ level.

To remedy this problem, we also use a hybrid approach that combines configuration interaction (CI) and all-order linearized
coupled-cluster methods \cite{SafronovaCIAll2009} and treat In as a trivalent system. This method allows us to consider
$5s5p^2$ configuration on the same footing with the $5s^2nl$ configurations and permits mixing of such levels. The main challenge in the theoretical treatment of systems with two or more valence electrons is the accurate treatment of both core-valence correlations and strong valence-valence correlations. In the CI+all-order method, the core-valence (and core-core) correlations are treated by the coupled-cluster all-order method, which is used to construct the effective Hamiltonian. The effective Hamiltonian is subsequently used in the configuration-interaction part of the method that treats the valence-valence correlations. The wave functions and the low-lying energy levels are determined by solving the multiparticle relativistic equation
$H_{\textrm{eff}}|\Psi\rangle = E|\Psi\rangle$. As a result, all of the correlation correction to the wave functions is treated at the all-order level. To improve the accuracy of the basis set for the orbitals of interest, we used an exact solution of the Dirac-Fock equations to obtain the $5p$, $5d$, $6s$, $6p$, $6d$, $7s$, $7p$, and $4f$ valence orbitals. The remaining orbitals are constructed using the B-splines, with the subsequent diagonalization of the combined basis.

The valence part of the polarizability is determined by solving the inhomogeneous equation of perturbation theory in the valence space, which is approximated as
\begin{equation}
(E_v - H_{\textrm{eff}})|\Psi(v,M^{\prime})\rangle = D_{\mathrm{eff},q} |\Psi_0(v,J,M)\rangle
\label{eq1}
\end{equation}
for a state $v$ with the total angular momentum $J$ and projection $M$ \cite{kozlov99a}. While $H_{\textrm{eff}}$ includes the all-order corrections as described above, the effective dipole operator $D_{\textrm{eff}}$ only includes random phase approximation (RPA) corrections at the present time. The CI+all-order method is generally used to extract properties of the low-lying states via Davidson's method which does not required full diagonalization of the matrix to solve the Schrodinger equation. While this allows precision determination of the $7p$ energies, numerical issues arise when calculating $7p$ polarizabilities. We find the iterative solutions of Eq.~\ref{eq1} do not converge in this case requiring full diagonalization of the matrix. Since it is exceptionally time-consuming to diagonalize very large matrixes used in the wave function calculation, we use the full calculation to sort the configurations in order of their importance. This allows us to reduce the matrix size for the direct solution of Eq.~\ref{eq1}. We correct for small numerical inaccuracy associated with the matrix truncation by recalculating dominant contributions to the polarizabilities using the matrix elements obtained with the full set of the configurations and experimental energies as described below. We note that these problems do not arise in the calculations of the low-lying state polarizabilities, such as $5p$, $6s$, and $6p$.

In the CC method, the polarizabilities are calculated using the sum-over-states approach.
 The valence contribution to scalar $\alpha_0$ and tensor $\alpha_2$ polarizabilities is evaluated as the sum over intermediate $k$ states allowed by the electric-dipole selection rules~\cite{MitSafCla10}
\begin{eqnarray}
    \alpha_{0}^v&=&\frac{2}{3(2j_v+1)}\sum_k\frac{{\left\langle k\left\|D\right\|v\right\rangle}^2)}{E_k-E_v}, \rm{~and} \label{eq-1} \nonumber \\
    \alpha_{2}^v(\omega)&=&-4C\sum_k(-1)^{j_v+j_k+1}
            \left\{
                    \begin{array}{ccc}
                    j_v & 1 & j_k \\
                    1 & j_v & 2 \\
                    \end{array}
            \right\} \nonumber \\
      & &\times \frac{{\left\langle
            k\left\|D\right\|v\right\rangle}^2}{
            E_k-E_v} \label{eq-pol},
\end{eqnarray}
             where $C$ is given by
\begin{equation}
            C =
                \left(\frac{5j_v(2j_v-1)}{6(j_v+1)(2j_v+1)(2j_v+3)}\right)^{1/2}. \nonumber
\end{equation}

\begin{table*}
\caption{\label{tab-7p1} Contributions to the $7p_{1/2}$ scalar polarizabilities of indium in $a_0^3$ calculated using the CC and CI+all-order approaches. Energy differences (in cm$^{-1}$) and absolute values of reduced matrix elements (in a.u.) are listed. Uncertainties are given in parentheses. }
\begin{ruledtabular}
\begin{tabular}{lrrrrr}
\multicolumn{1}{c}{Contr.} &
\multicolumn{1}{c}{$\Delta E_{\rm{expt}}$} &
\multicolumn{2}{c}{Matrix elements} &
\multicolumn{1}{c}{$\alpha_{0}$} &
\multicolumn{1}{c}{$\alpha_{0}$} \\
\multicolumn{2}{c}{} &
\multicolumn{1}{c}{CC} &
\multicolumn{1}{c}{CI+all} &
\multicolumn{1}{c}{CC} &
\multicolumn{1}{c}{CI+all}  \\
[0.3pc]\hline
$6s$&  -14488.5&   0.683(63)& 0.649(18) &    -2(0)   & -2(0) \\
$7s$&   -2559.6&  12.215(57)&12.21(20) &-4264(40)   & -4258(130) \\
$8s$&    1775.6&  12.301(51)&12.335(42) &  6235(52)  & 6269(43) \\
$9s$&    3857.6&   2.252(22)&       &   96(2)    &  \\         [0.4pc]
$ 5d_{3/2}$&   -5969.2&   2.179(30)& 1.941(90) &  -58(1)    &  -46(4)     \\
$6d_{3/2}$&      187.1&  21.50(27) &21.24(20)  &180679(4500)&  176471(3300)\\
$ 7d_{3/2}$&    2975.0&  10.8(1.2) &       &   2885(660)&   \\
$ 8d_{3/2}$&    4474.5&   4.670(45)&       &    356(68) &  \\   [0.4pc]
$5s 5p^2$       &         &            &       &      -5    & -5\\
 Other          &         &            &       &     309(62)       & 3821(80)  \\
Core            &         &            &       &        30  & 3  \\
Total           &         &            &       &186259(4600)&  182253(3300) \\
\end{tabular}
\end{ruledtabular}
\end{table*}

\begin{table*}
\caption{\label{tab-7p2} Contributions to the $7p_{3/2}$ scalar and tensor polarizabilities of indium in $a_0^3$ calculated using the CC and CI+all-order approaches. Energy differences (in cm$^{-1}$) and absolute values of reduced matrix elements (in a.u.) are listed. Uncertainties are given in parentheses. }
\begin{ruledtabular}
\begin{tabular}{lrrrrrrr}
\multicolumn{1}{c}{Contr.} &
\multicolumn{1}{c}{$\Delta E_{\rm{expt}}$} &
\multicolumn{2}{c}{Matrix elements} &
\multicolumn{1}{c}{$\alpha_{0}$} &
\multicolumn{1}{c}{$\alpha_{0}$} &
\multicolumn{1}{c}{$\alpha_{2}$} &
\multicolumn{1}{c}{$\alpha_{2}$}  \\
\multicolumn{2}{c}{} &
\multicolumn{1}{c}{CC} &
\multicolumn{1}{c}{CI+all} &
\multicolumn{1}{c}{CC} &
\multicolumn{1}{c}{CI+all} &
\multicolumn{1}{c}{CC} &
\multicolumn{1}{c}{CI+all}  \\
[0.3pc]\hline
$6s_{1/2}$& -14599.9 & 1.131(87)&1.086  &    -3(0)    &-3.0 &      3(0) & 3.0 \\
$7s_{1/2}$& -2671.0  &16.933(91)&16.922 &-3927(42)    &-3921 &  3927(42) &3921 \\
$8s_{1/2}$&  1664.1  &18.280(87)&18.319 & 7345(71)    &7377 &  -7345(71)& -7377 \\
$9s_{1/2}$&  3746.1  & 3.081(33)&       &  93(2)      &       &     -93(2)&  \\      [0.4pc]
 $5d_{3/2}$& -6080.7 & 0.769(47)&0.755  &-4(0)        & -3.4&    -3(0)  &  -3  \\
$6d_{3/2}$&      75.6& 9.60(12) &9.473  &44566(1100)  & 43403& 35652(910)& 34722  \\
$7d_{3/2}$&    2863.5& 5.27(57) &       & 355(76)     &      &   284(61) &       \\
$8d_{3/2}$&    4363.0& 2.20(21) &       & 41(8)       &      &    33(6)  &        \\ [0.4pc]
$5d_{5/2}$&   -6057.4& 2.346(94)&2.308 &    -33(3)   &-32 &      7(1) &    6 \\
$6d_{5/2}$&     125.5& 29.07(80)&28.46 &246375(13600)&236142 &-49275(2700)&-47229      \\
$7d_{5/2}$&    2889.1& 15.4(2.5)&      &   2986(970) & &  -597(200)&      \\
$8d_{5/2}$&    4382.2&  6.48(82)&      &    350(88)  & &   -70(18) &       \\
$5s 5p^2$  &          &          &      & -280       &  -280           &56 &   56        \\
Other     &          &          &      &    323(64)  &4650 & -66(13)   & -333  \\
Core      &          &          &      &  29.6       & 3&      0    &     0         \\
Total     &          &          &       &298215(13600)&287332 &-17488(2870)& -16233  \\
\end{tabular}
\end{ruledtabular}
\end{table*}
The contributions to the $7p_{1/2}$ and $7p_{3/2}$ polarizabilities in indium (in units of $a_0^3$) are given in Tables~\ref{tab-7p1} and \ref{tab-7p2}. The $\Delta E=E_k-E_v$ energy difference calculated using the experimental values \cite{NIST} and the absolute values of the reduced matrix elements obtained using both CC and CI+all-order methods are also listed.  The uncertainty of the CC matrix elements is estimated using the method described in \cite{Safronova2013}. Briefly, four different CC calculations are carried out, including two \textit{ab initio} calculations with and without the inclusion of the partial triple contributions and two corresponding calculations where higher excitations are estimated using a scaling procedure. The maximum differences of the final values and the other results provide the uncertainty estimates. To estimate the uncertainty in the CI+all-order matrix elements, we carry out a calculation that combines configuration interaction and the second-order many-body perturbation theory (CI+MBPT) \cite{DzuFlaKoz96}. In this method, the effective Hamiltonian is constructed using the second-order MBPT rather than the all-order coupled-cluster approach, omitting all higher-order core-valence and core-core corrections. The differences of the CI+all-order and CI+MBPT results provide a rough estimate of the uncertainties. We could not use this approach for the $7p_{3/2}-6d_{5/2}$ matrix element since the CI+MBPT method places the $5s5p^2$ configuration very close to the $6d_{5/2}$ level, resulting in incorrect level mixing; therefore we do not quote uncertainties for the $7p_{3/2}$ polarizability. We estimate the uncertainty of the CI+all-order $7p_{3/2}$ scalar polarizability value at 2\% based on the $7p_{1/2}$ uncertainty.  Finally, we estimate the uncertainty of the $7p_{3/2}$ tensor polarizability to be 10\% by considering the uncertainties in the analogous calculation for the $6p_{3/2}$ state, while also recognizing that, for the $7p$  case, there is significantly more severe cancellation of terms in the relevant sum. This work provides an excellent test of these methods to evaluate theoretical uncertainties in CC and CI+all-order frameworks, which is crucial for many other applications where experimental data are not available but uncertainty estimates are required. The relative uncertainty in the polarizability contribution is twice the relative uncertainty of the matrix element. The remaining valence contributions, not explicitly listed in the tables are grouped together in the rows labelled ``Other''.  The contribution of the ionic core calculated in the RPA approximation as described in \cite{Safronova2013} is listed in the row labeled ``Core". It is negligible for the $7p$ states.

\begin{table*}
\caption{\label{tab} Comparison of experimental and theoretical results for In polarizabilities. $\Delta \alpha_{0}$ in the third column of results refers to the $6s - 5p_{1/2}$ polarizability difference. CC $6s$, $5p_{1/2}$, and $6p$ values are from \cite{Safronova2013}.
(a) Ref. \cite{Indium2013}. (b) Ref. \cite{Indium2016}.}
\begin{ruledtabular}
\begin{tabular}{lccccccccc}
\multicolumn{1}{c}{Method} &
\multicolumn{1}{c}{$\alpha_0(6s)$} &
\multicolumn{1}{c}{$\alpha_0(5p_{1/2})$} &
\multicolumn{1}{c}{$\Delta \alpha_{0}$} &
\multicolumn{1}{c}{$\alpha_0(6p_{1/2})$} &
\multicolumn{1}{c}{$\alpha_0(6p_{3/2})$} &
\multicolumn{1}{c}{$\alpha_2(6p_{3/2})$} &
\multicolumn{1}{c}{$\alpha_0(7p_{1/2})$} &
\multicolumn{1}{c}{$\alpha_0(7p_{3/2})$} &
\multicolumn{1}{c}{$\alpha_2(7p_{3/2})$} \\
\hline
CC            & 1056(27)&61.5(5.6)&995(28)&   7817(155) &10506(180)& -1432(42)& 1.863(46)$\times10^5$&2.98(14)$\times10^5$&-1.75(29)$\times10^4$  \\
CI+all  & 1055(7)          & 62.5(2.0)          &       992(7)   &7630(120) & 10259(230) & -1407(40) &1.823(33)$\times10^5$&2.87(6)$\times10^5$&-1.62(16)$\times10^4$\\
\hline
Expt.    & 1050(6)              &               & 988.0(2.7)$^{a}$&7590(37)$^{b}$   &       &       &    1.811(04)$\times10^5$           &    2.876(06)$\times10^5$          & -1.43(18)$\times10^4$ \\
\end{tabular}
\end{ruledtabular}
\end{table*}

While the calculation of the CI+all-order polarizabilities does not involve the sum-over-states expressions (Eq. \ref{eq-1}), it is instructive to extract several low-lying contributions using the above expression. As noted below, we replace these with the more accurate  experimental energies and CI+all-order matrix elements obtained in the full-scale computation. The differences are well below the expected accuracy of the calculations with the exception of the $7p-6d$ contribution which has a very small energy difference. While the CI+all-order method reproduces the energy levels with about 0.5\% precision, even a 50~cm$^{-1}$ error in the $7p-6d$ theoretical energy difference very strongly affects polarizabilities, and the experimental interval must be used. As expected, the $7p-6d$ contribution is strongly dominant, giving 97\% for both $7p$ scalar polarizabilities. We have also calculated the $6s$, $5p$, and $6p$ polarizabilities and estimated their uncertainties using an improved basis set, added the Breit interaction and improved constriction of the configuration space in comparison with the 2013 work \cite{Safronova2013}.

\section{Discussion and Comparison of Results}

All new and updated theoretical results are summarized alongside corresponding experimental values in Table~\ref{tab}. As noted earlier, while measured Stark shift constants reflect the differential shift between the $6s$ and $7p$ states, the $6s$ Stark shift is nearly two orders of magnitude smaller, and has been measured with high accuracy previously in our group \cite{Indium2013}. Therefore, it is straightforward, without loss of accuracy, to convert our results to polarizabilities of the $7p$ states themselves. Table~\ref{tab} summarizes these results, in atomic units, and also includes older experimental measurements from 2013 and 2016. We note that for the case of these older results, a small error was discovered in the numerical factor used to convert the measured Stark shift constants to polarizability in atomic units. The corrected polarizability numbers for the $6s$ and $6p_{1/2}$ states in atomic units are included in Table~\ref{tab} along with our new results in the final row of the table. Revised CI+all-order values for the $6s$ and $6p$ polarizabilities are in better agreement with the CC values, resolving a previous discrepancy. Considering all excited state experimental and theory information now available, we find that the central values obtained with the CI+all-order method are in significantly better agreement with the experiment for the $6p$ and $7p$ polarizabilities, which is likely due to direct inclusion of the three-particle configurations beyond the $5s^2nl$. We find that such configurations contribute a few percent via the level mixing to the $5s^2nl$ wave functions. In every comparison, the experimental values agree to within 0.5\% with the quoted CI+all-order theoretical predictions.

Because of the large matrix elements and very small energy differences associated with the $7p-6d$ terms within the infinite sums that make up the polarizabilities, we can make a straightforward determination of these particular matrix elements using our experimental values and the known energy splittings. By subtracting from our experimental polarizability value the residual terms of the theoretical sum (which in total represent only a few percent of the net polarizability), we can isolate the dominant term in the sum, and then compute a recommended value for the particular matrix element of interest. This procedure does not lead to any additional uncertainties, as the error in the residual terms of the infinite sum are very small compared to the experimental uncertainty. A similar procedure was undertaken for the case of the $6p-5d$ matrix element in \cite{Indium2016}. In the present work, we infer the following values for two indium reduced matrix elements (in atomic units):  
\begin{eqnarray}
   \langle 6d_{3/2} || D || 7p_{1/2} \rangle = 21.17(04) \hspace{1mm} \nonumber \\
   \langle 6d_{5/2} || D || 7p_{3/2} \rangle = 28.49(11).  \nonumber
\end{eqnarray}
When we compare these values to the relevant theoretical entries in Tables \ref{tab-7p1} and \ref{tab-7p2} we see particularly good agreement with the corresponding CI+all entries there.

%

\section{Conclusion}

We have completed new high-precision measurements of polarizabilities in the highly excited $7p$ states of $^{115}$In, and in parallel developed \textit{ab initio} theoretical calculations of those same quantities. The measurements are the first of their kind in highly excited states of trivalent group IIIA atoms. Similarly, this work represents the first calculation of polarizabilities of such high excited states with the CI+all-order method, which was initially designed to provide values for low-lying states. A number of difficulties were overcome to adapt the theoretical approach for this task. By combining the present polarizability measurements with recent ones in lower-lying states of indium, we have directly demonstrated the value of such experimental benchmarks in guiding theoretical work forward via their ability to discern between competing models. The experimental values for the $6p$ and $7p$ states are clearly in better agreement with the CI+all-order calculations that treat In as a three-electron system, demonstrating the importance of the configuration mixing. The comparison also validates the procedures for the evaluation of theoretical uncertainties in both approaches. Such work is essential to allow the continued development of theory necessary for robust tests of fundamental physics in these trivalent systems. Future experimental work will extend these two-step measurements of excited state polarizabilities to the thallium system, where a similarly detailed comparison of experiment and theory should be possible in this heavier trivalent system.

\section*{Acknowledgments}
We thank Mikhail Kozlov and Sergey Porsev for helpful discussions. The work of M.S.S. was supported in part by the National Science Foundation grant PHY-1620687. The experimental work described here was completed with the support of the National Science Foundation RUI program, through Grant No. PHY-1404206.

\bibliography{IndiumABU2017,theory}

\end{document}